# CONSOLIDATED MODEL OF PROCEDURES FOR WORKFLOW MANAGEMENT

Stanka Hadzhikoleva and Emil Hadzhikolev

Department of Mathematics and Computer Science, University of Plovdiv, Bulgaria
stankah@uni-plovdiv.bg and hadjikolev@uni-plovdiv.bg

*ABSTRACT*

*The article presents an approach to automation of business processes by means of a consolidated model describing a class of processes. Rules and examples for building a consolidated model are given. The model is validated through development of a software application called COMPASS-P for monitoring of procedures for evaluation and accreditation of education.*

*KEYWORDS*

*consolidated model of procedures, generalized model of a class of procedures, workflow model, evaluation of education, evaluation of the quality in education.*

## 1. INTRODUCTION

There are many different business processes in different organizations that occur in a similar way, involve numerous similar activities or tasks performed by employees with different rights and responsibilities. A typical example is the management of documents between the different departments in large companies, in public administration, institutions working in the field of healthcare, education, etc. The activities themselves usually involve entering information about occurred or upcoming events, received or approved documents and so on.

The reason for conducting the survey was the task for automation of monitoring activities related to procedures for evaluation and accreditation in higher education carried out by the Bulgarian National Evaluation and Accreditation Agency (NEAA) [8]. The analysis showed *significant similarities* between the procedures:

- *the same activities carried out by the same users* – for instance, an accountant enters information about a sum of money received and the remittance date; a secretary of a standing committee by area of higher education (SCAHE) or a secretary of accreditation council (AC) enters information about a protocol with a proposal for expert group (EG) election, including protocol number and date, etc.;
- *strictly defined sequence of activities in time* – for instance, the SCAHE decision for opening a procedure is always registered first, then payment is made by the university, an EG is appointed and so on.

*The differences* came from the specifics of the separate procedure types, for instance:

- some procedures are managed by SCAHE and others by AC;
- some procedures involve appointing an EG whereas with others the need of an EG is decided during the execution of the procedure;
- some procedures require the university's opinion regarding the result of the procedure, others do not require this opinion, etc.

A question arose: "***Is it possible to create a generalized model for all procedures***, a model that would allow the use of a uniform model of the data and their unified processing?"

## 2. AUTOMATION OF BUSINESS PROCESSES

There are two basic approaches to *business processes management*:

- **activity oriented** – related to the processing of complex nonlinear processes; solves different tasks, independent of each other; the tasks implementation involves certain deadlines;
- **workflow oriented** – the processes occur linearly, the implementation of a step depends on parameters obtained during the execution of previous steps.

The presented survey is focused on the *workflow-oriented business processes*.

The automation of workflows can be organized in different ways. One possibility is to *use a software system for business processes management*. There are many applications for workflow modeling. Most of them support in various degrees the Business Process Model and Notation (BPMN) 2.0 standard [10] supported by OMG [9]. The development of the standard is coordinated with numerous leading companies such as IBM, Oracle, SAP and others, which ensures its importance. Some software systems also provide features to manage processes based on the established models, for instance open-source engines such as: jBPM [6], Activiti [1], Bonita Open Solution [3], the paid suite ActiveVOS [2] and so on. These applications are suitable for modeling activity-oriented, heterogeneous business processes, they require specialized maintenance, they are often difficult to configure and support multiple functionalities that are not used in the different specific cases of application.

The other way to automate workflows is by *building a software application* specialized in the specific subject area. *Reference process models* can be used in this case. There are different libraries with ready-made reference process models created by different organizations such as SAP Reference Model [4], Supply Chain Operations Reference (SCOR) [12] and others.

The article offers an approach for *building a software application for management of business processes using a consolidated model* for description of a class of similar processes.

## 3. CONSOLIDATED MODEL OF PROCESSES

Marcello La Rosa [7] uses the term "configurable process model" to describe a model that presents a consolidated view of a class of business processes. He is referring to a complex parameterized process model with many perspectives (variants) of execution. The purpose is to help analysts during the configuration phase and guarantee the accuracy of the resulting process models. La Rosa suggests a notation for modeling configurable business processes, called C-iEPC. It is an extension of the C-EPC notation [11] and supports mechanisms for presenting a range of options in different variant points of the process such as inclusion of resources, data and physical objects involved in the execution of the tasks, etc.

We use the term "consolidated model" (CM) in a different sense. *The consolidated model of processes is a generalized model containing models of many different, more elementary processes.* The creation of a CM is a process which creates a generalized process model of numerous individual processes. Each of the elementary processes can be executed independently on the consolidated model.

In other words, a **CM** of the processes **describes all types of processes and the various options for their execution. It contains all the steps of all processes.** The description of each step includes numerous characteristics, among which **conditions for the implementation of the step** and **output data** (which initiate/change the parameters that control the processes). Which step will be executed in a given specific process and whether it will be executed depends on the conditions for implementation of the individual steps.

*The use of a CM is appropriate if any or all of the following conditions are available:*

- numerous processes that occur in a similar way – as a sequence of steps and logic of implementation;
- the processes have many common steps (i.e. steps that occur in all processes);
- it is necessary to create a software system that supports uniform processing of the data collected during the execution of the processes.

## 4. CREATING A CONSOLIDATED MODEL OF PROCESSES

When creating a CM of processes it is recommended that *the existing elementary processes upon which a CM is built belong to the same subject area and describe activities of the same type* differing only in few of the details. It is practically possible but inappropriate to create a CM for elementary processes from different subject areas, having no common activities.

A CM have two forms of presentation:

- **graph** – hierarchical, standard form;
- **linear form**.

The creation of a CM of processes includes the following activities:

1. *Studying and analyzing all processes and specifying their steps.* A *step* includes activities performed by one person that can be implemented consecutively, without requiring intervention, conducting an activity or a result by another person or a software component. The step is the smallest indivisible unit that builds a process model.
2. *Determining the common steps of the processes.* A *common step* is a step involved in at least 2 process models. The common steps are noted for each separate process. If the processes share many common steps, construction of a consolidated model can be attempted. If the number of common steps is small, the designer is to decide whether the use of a CM is appropriate.
3. *The development of a CM of processes* is subject to the following rules:
    **3.1** A CM has one initial step and one final step. If the initial and the final steps of the processes are not the same, common initial and final steps are introduced artificially.
    **3.2** In a CM the common steps are entered first. They must be in the correct sequence, i.e. in the order in which they are implemented in all the processes.
    **3.3** Between any two adjacent steps in a CM the steps of the elementary processes are included in different ways according to the presentation:
      **3.3.1.** In a "graph" form – in separate branches through "OR-connection".
      **3.3.2.** In a linear form – the steps of each specific process are arranged consecutively, for instance in the following way: the specific steps of the first process are entered into the CM, then the steps of the second process are added and so on. The intermediate steps between two common steps of the CM must be in the correct sequence for each and every process. The arrangement in total of the intermediate steps in a CM does not matter, i.e. it is possible to alternate steps belonging to different processes. What is important is to set the steps of each process in the correct sequence.
    **3.4** For each step the conditions under which it can be executed are described in a CM. This may involve introduction of specific parameters: elementary process type, time parameters, etc.

Fig. 1 shows two processes: A and B. The common steps are marked in grey and labeled C1, C2 and C3. In third position a CM in standard form is shown. Four possible options for constructing a linear form of a CM are given to the right. CM 1 and CM 2 are correct because they observe the sequence of the common steps of the processes, as well as the sequence of the specific steps of the processes (rule 3.3.2). In model 2, the specific steps of the processes between the common steps C1 and C2 are not arranged in the correct sequence of the processes (i.e. all steps of one process first, then all steps of the other process), but the arrangement of implementation of the steps in the processes is observed (i.e. the steps of each process themselves are in the correct sequence), thus CM 2 is correct. CM 3 is incorrect because A4 comes after A5, i.e. the sequence of the specific steps of process A in the CM is not observed (rule 3.3.2). CM 4 is also correct – it pays attention to the fact that step A3 is alternative to the common step A2(C2). A3 must be located between C1 and C3 (rule 3.3.2), which is observed in CM 1, CM 2 and CM 4.

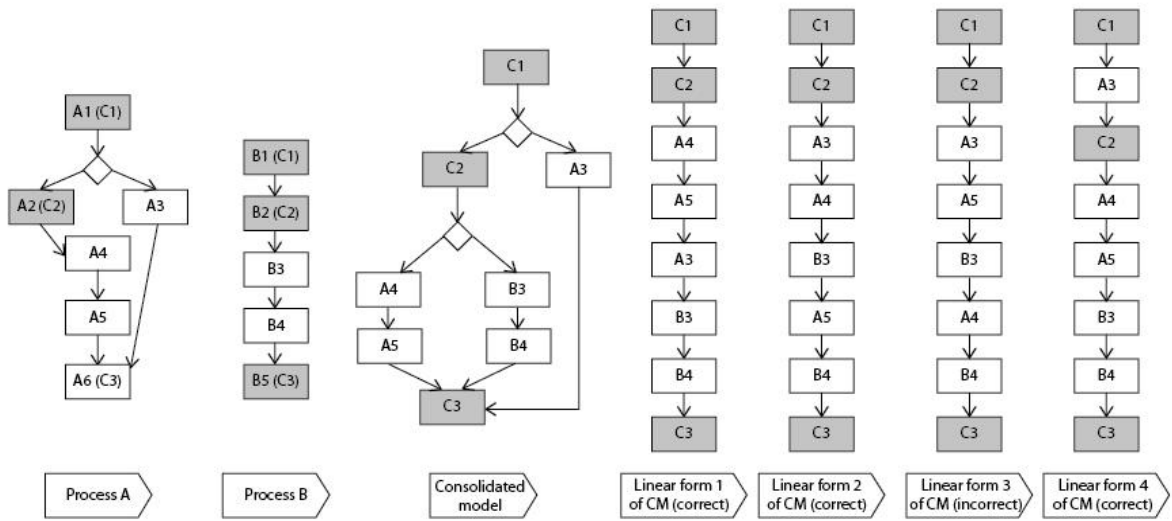

Figure 1. Example for development of a consolidated model

One standard process model can be presented in several linear forms. Specific forms do not influence the performance of the processes which are executed in the same way in all variants. In a standard form it is easy to see the operations connecting the steps, whereas in a linear form of presentation, the steps are arranged sequentially. Conditions for implementation are set with both approaches. With the linear approach operations are included in the conditions for implementation of a step.

## 5. SOFTWARE MODELING

The main task in developing a software system using a CM is the creation of an appropriate model of a procedure and a step. The model proposed by us has been used for building an application for management of procedures for quality evaluation but can be extended and adapted for other areas as well.

The model of objects and the connection between them is shown in Fig. 2. The basic object is the *Step*. The step is the main building block of the procedures and is described by the following major characteristics:

- **number** - indicates the sequence of implementation of the step in a linear form of CM;
- **types of procedures** to which the step belongs (i.e. procedures in which the step can be executed);
- **a list of fields** involved in the step;
- **input-output data** – results of the step performance that influence the further development of the procedure;
- **specific data** is information processed by the step;
- **editing rights** – indicate user roles that have rights to edit the step;
- **viewing rights** – indicate user roles that have rights to view the step;
- **editability** – indicates whether the step can be edited;

- **visibility** – controls the visibility of the step in the procedures;
- **conditions for implementation** is a logical expression describing the requirements that must be satisfied to perform the step. These are typically values of parameters initialized/modified by other steps and set for specific types of procedures. They can be mandatory or optional;
- **visualization** - presents the step in view mode or edit mode;
- **editing** – determines whether the step can be edited depending on different variables – role, procedure type, visibility, editability, etc.;
- **conditions for completion** specify the conditions under which the step is considered completed – for instance, mandatory data introduced, expiry of a certain period of time (i.e. period of time during which the step must be performed, after the expiry the step is skipped and the next step in the procedure becomes ongoing) and so on.

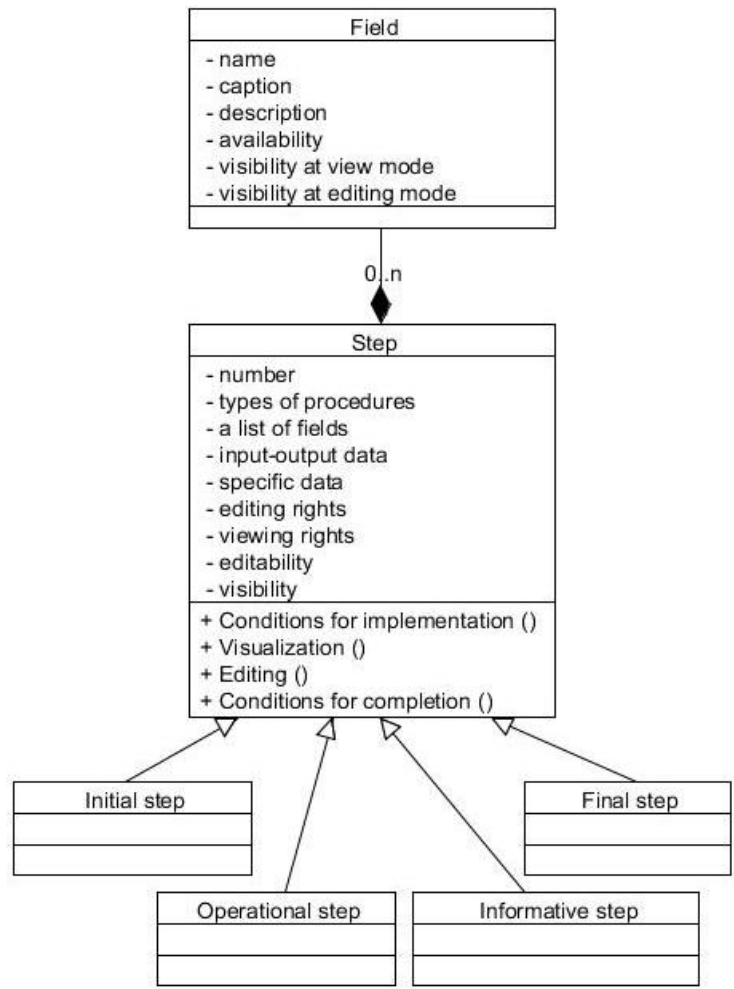

Figure 2. Model of a step

*The step consists of fields* the major characteristics of which are:
- field **name**;
- **caption** with information about the field, visible to users;
- field **description**;
- **availability** - determines whether entering value in the field is mandatory for the completion of the step;

- **visibility at "view" mode** – indicates whether the field is displayed when the step is shown in view mode;
- **visibility at "editing" mode** – indicates whether the field is displayed when the step is shown in edit mode, and so on.

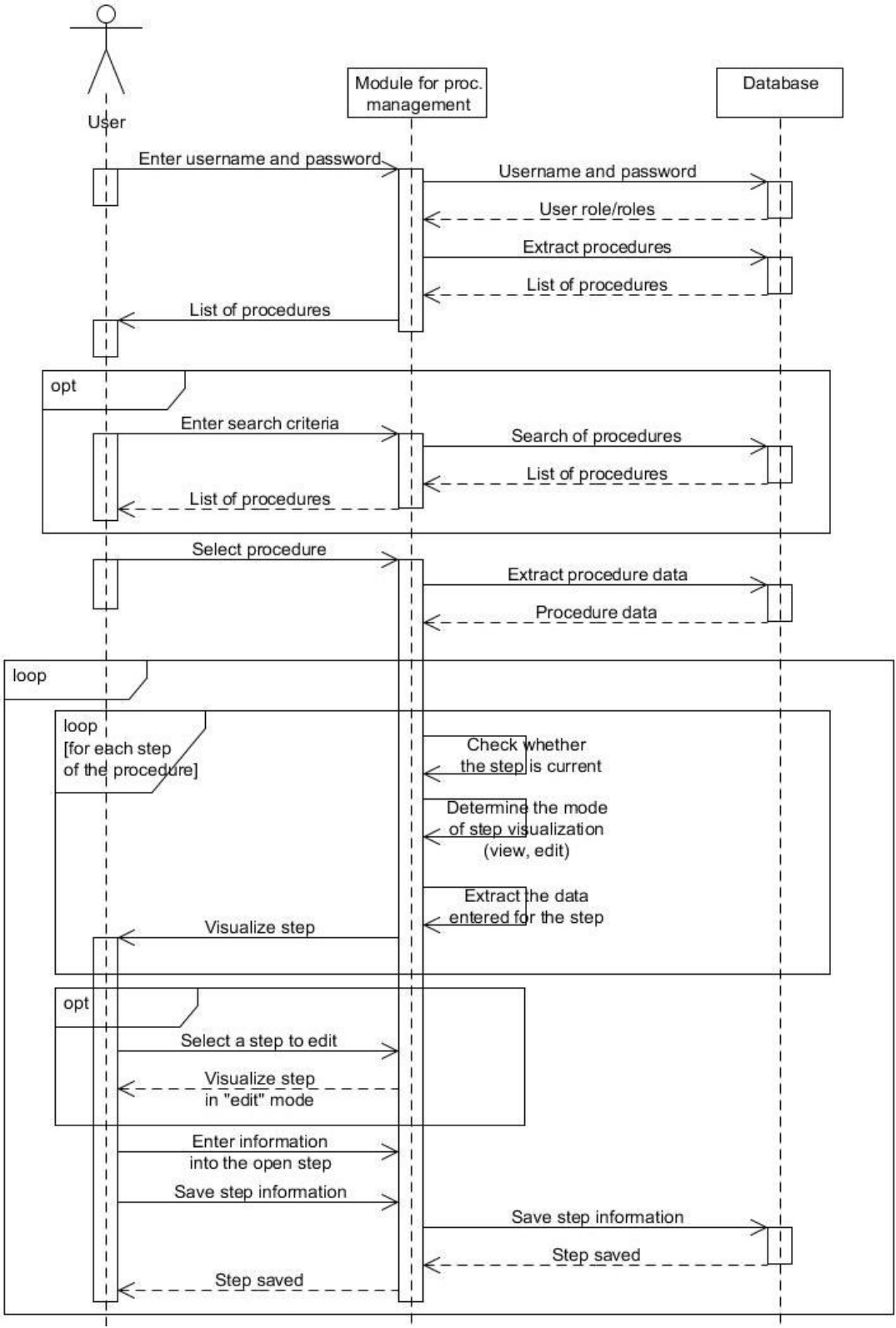

Figure 3. Editing a procedure

*The management of a CM of processes* includes methods for:

- **determination of current step**. One approach for determining the current step in a linear form of a CM is a consecutive scan of all the steps and selection of the one which has blank mandatory fields and the conditions for implementation of which are met;
- **procedure visualization** – visualizes a process instance, the current step being shown in "edit" mode and the others – in "view" mode;
- **step management** includes activities related to:
    - verification of the conditions for step performance;
    - preparation of specific and input data;
    - setting visualization mode – view, edit;
    - saving the state, etc.
- **procedure management** uses the activities "procedure visualization", "determination of current step", "selection", "editing", "saving the state" and so on.

*Editing a procedure* by a user involves procedure visualization and editing of one or more steps. After each step edit the procedure management module re-determines the current step and its view, depending on the user role (roles) (fig. 3).

## 6. EXPERIMENTS

The proposed concept for a CM describing a class of procedures is validated by building a specialized software application COMPASS-P for management and monitoring of procedures for evaluation of quality in education. The main requirements for the application are: *to register the progress of procedures* (without managing the document flow; only by citing the relevant documents), *to register data entered by employees with different responsibilities* (after authorization), *to monitor compliance with the terms of the procedures* (in accordance with the established regulations), *to provide free access to the collected information* for all users of the system, *to provide tools for search by many different criteria and generation of reports*.

Thirteen types of accreditation procedures of the NEAA have been analyzed and specified and fifteen common steps have been derived from them. Six roles have been defined with different rights and access to the procedures – administrator, accountant, AC secretary, SCAHE secretary, clerk and observer. A CM of the procedures has been created. Each step of the CM requires entering information about a certain event (a meeting held, a payment made, a statement received, a notification sent and so on) by a user with a certain role. The system users have different access to the steps (view or edit) depending on their role, the steps' status (completed, current, future) and the procedures' status (current, archived). COMPASS-P has been tested in the NEAA with real data. Detailed information about the application is presented in [5].

## 7. CONCLUSIONS AND FUTURE WORK

The article presents an approach to managing work processes that occur in a similar way (as a sequence of steps and logic of implementation) and have many common steps (i.e. steps that occur in all processes). Rules are proposed for building a consolidated model describing a class of processes. The use of a CM has numerous advantages:

- it is not necessary to model and implement numerous similar procedures;
- procedures can be processed in the same way, for instance for visualization, search by different criteria, generation of reports and so on;
- different summaries for all types of procedures can be easily made due to the opportunity to treat them as a single procedure;
- when the underlying work model is changed, the implementation is modified once only and so on.

The possibility for visual process modeling is of major importance for widespread use. In the current version of COMPASS-P we use a specific instance with description of the process. Subject of future work is the development of a formal, parameterized model for describing processes/procedures aimed at automated design of CM of procedures. Another important task is the creation of algorithms to verify the correctness of a consolidated model.

## ACKNOWLEDGEMENTS

This work is part of projects BG051PO001-4.3.04-0064 "Plovdiv electronic University (PeU): National Standard for Implementation of Quality e-learning in the System of Higher Education" and BG051PO001-3.1.08-0041 "Standardization and Integration of Heterogeneous Information and management University Systems (SIRIUS)", funded by the "Human Resources Development" OP at the ESF.